\documentclass[letterpaper]{article}
\usepackage[preprint]{aaai2027}
\usepackage[hyphens]{url} 
\usepackage{graphicx} 
\usepackage{natbib} 
\usepackage{caption} 
\usepackage{amsmath}
\usepackage{amssymb}
\usepackage{booktabs}
\usepackage{xcolor}
\usepackage{tikz}
\usetikzlibrary{arrows.meta,calc,fit,positioning}

\definecolor{specTeal}{HTML}{176B68}
\definecolor{specTealLight}{HTML}{DCEFED}
\definecolor{specBlue}{HTML}{315C8C}
\definecolor{specBlueLight}{HTML}{DCE7F2}
\definecolor{specRed}{HTML}{A4473E}
\definecolor{specRedLight}{HTML}{F4DFDC}
\definecolor{specGold}{HTML}{B77A1F}
\definecolor{specGoldLight}{HTML}{F5E9CF}
\definecolor{specGray}{HTML}{5D6670}
\definecolor{specGrayLight}{HTML}{EDF0F2}

\newcommand{\benchmark}{\textsc{SpecPath}}
\newcommand{\numbasetasks}{5}
\newcommand{\numfamilies}{5}
\newcommand{\numhistories}{5}
\newcommand{\numconditions}{7}
\newcommand{\numrepos}{5}
\newcommand{\numendpoints}{7}
\newcommand{\numscaffolds}{2}
\newcommand{\numsystems}{14}
\newcommand{\numruns}{3}
\newcommand{\numplanned}{1,470}
\newcommand{\numscored}{1,194}
\newcommand{\directfcr}{78.8\%}
\newcommand{\pathfcr}{78.7\%}
\newcommand{\conditionalviolations}{35 of 100}

\title{SpecPath: Testing Coding Agents Across Contract-Equivalent
Specification Histories}
\author{
Yangfan Wu\textsuperscript{\rm 1},
Haozhe Wang\textsuperscript{\rm 1},
Huanyu Yang\textsuperscript{\rm 2},
Jianmin Ji\textsuperscript{\rm 2},
Fangzhen Lin\textsuperscript{\rm 1}
}

\affiliations{
\textsuperscript{\rm 1}The Hong Kong University of Science and Technology\\
\textsuperscript{\rm 2}University of Science and Technology of China
}

\begin{document}

\maketitle

\begin{abstract}
Modern coding agents increasingly appear capable of following complex software
requirements, yet their success leaves a critical ambiguity: do they resolve
the active specification, or merely follow the most salient path by which it
was stated? We identify \emph{specification-path sensitivity}, a failure mode in
which requirement histories that are equivalent in their final meaning lead the
same agent system to produce behaviorally different programs. This reframes
evolving-requirement evaluation as active-contract resolution: before writing
code, an agent must determine which requirements still count.

Building on this view, we introduce \benchmark, a diagnostic evaluation that
holds the repository, final contract, verifier, agent system, and execution
budget fixed while changing only the revision path that leads to the contract.
Rather than treating each patch as an isolated pass or failure, \benchmark\ uses
paired executable outcomes to reveal whether an agent realizes the same tested
behavior across contract-equivalent histories. Across five calibrated software
tasks and fourteen coding-agent configurations, aggregate direct and
revision-history accuracy is nearly unchanged; nevertheless, 35 of 100 complete
blocks that succeed on the direct specification fail on at least one equivalent
history. These results show that implementation success on a consolidated
request does not guarantee specification-path invariance. Evaluating evolving
requirements therefore calls for controlled tests of whether agents are robust
to the path by which a specification becomes final.
\end{abstract}

\providecommand{\benchmark}{\textsc{SpecPath}}
\providecommand{\numbasetasks}{5}
\providecommand{\numfamilies}{5}
\providecommand{\numhistories}{5}
\providecommand{\numconditions}{7}
\providecommand{\numrepos}{5}
\providecommand{\numendpoints}{7}
\providecommand{\numscaffolds}{2}
\providecommand{\numsystems}{14}
\providecommand{\numruns}{3}
\providecommand{\numplanned}{1,470}
\providecommand{\numscored}{1,194}
\providecommand{\directfcr}{78.8\%}
\providecommand{\pathfcr}{78.7\%}
\providecommand{\conditionalviolations}{35 of 100}

\section{Introduction}
\label{sec:introduction}

Software requirements rarely remain consolidated while an agent works. Issue
updates and review comments may replace a rule, withdraw a workaround, or
narrow its scope. A reliable coding agent must therefore determine not only
\emph{how} to implement a request, but \emph{which} requests still form the
active contract. The second problem logically precedes the first: even flawless
implementation skill realizes the wrong program if the agent resolves the wrong
contract.

The source of this difficulty is a mismatch between representation and state.
A conversation is an append-only record, but its specification is mutable. When
turns are additive, accumulating them is correct. When a later turn supersedes,
scopes, or cancels an earlier one, neither ``remember everything'' nor ``follow
the latest turn'' is sufficient. The agent must instead recover a normalized
set of currently binding obligations from the complete history. We call this
inference step \emph{active-contract resolution}.

\begin{figure*}[t]
    \centering
    \resizebox{0.98\textwidth}{!}{\begin{tikzpicture}[
  x=1cm,y=1cm,
  >={Latex[length=2.0mm,width=1.3mm]},
  flow/.style={->,line width=.9pt,draw=specBlue!80},
  history/.style={rounded corners=2pt,draw=specGray!55,line width=.55pt,
    fill=white,align=left,inner sep=5pt,font=\scriptsize},
  active/.style={history,draw=specTeal,line width=.75pt,fill=specTealLight},
  obsolete/.style={history,draw=specRed!75,dashed,line width=.7pt,
    fill=specRedLight,text=specRed},
  resolve/.style={rounded corners=2pt,draw=specBlue!65,line width=.6pt,
    fill=white,align=center,inner sep=4pt,font=\scriptsize},
  run/.style={rounded corners=2pt,draw=specGray!55,line width=.6pt,
    fill=white,align=left,inner sep=5pt,font=\scriptsize},
  result/.style={rounded corners=2pt,draw=specGray!55,line width=.6pt,
    fill=white,align=left,inner sep=5pt,font=\scriptsize},
  proposition/.style={align=center,font=\scriptsize\bfseries,text=specGray},
  lane/.style={anchor=east,font=\scriptsize\bfseries,text=specGray}
]
  \node[proposition] (p1) at (2.25,5.15)
    {DIFFERENT HISTORIES\\[-1pt]\normalfont\(H_D \ne H_O\)};
  \node[proposition,text=specBlue] (p2) at (6.35,5.15)
    {SAME ACTIVE CONTRACT\\[-1pt]\normalfont
     \(\operatorname{Resolve}(H_D)=\operatorname{Resolve}(H_O)=C^\star\)};
  \node[proposition] (p3) at (10.35,5.15)
    {MATCHED EXECUTION\\[-1pt]\normalfont fresh runs; controls fixed};
  \node[proposition,text=specRed] (p4) at (14.85,5.15)
    {DIFFERENT BEHAVIOR\\[-1pt]\normalfont \(B_D \ne B_O\)};
  \draw[->,specGray!55,line width=.7pt] (p1.east)--(p2.west);
  \draw[->,specGray!55,line width=.7pt] (p2.east)--(p3.west);
  \draw[->,specGray!55,line width=.7pt] (p3.east)--(p4.west);

  \node[lane] at (.55,3.35) {Direct};
  \node[lane] at (.55,1.20) {Override};

  \node[active,text width=3.05cm,anchor=west] (direct) at (.75,3.35)
    {\textbf{State \(C^\star\) directly}\\[2pt]
     required missing \(\rightarrow\) raise\\
     optional missing \(\rightarrow\) default};

  \node[obsolete,text width=1.20cm,anchor=west] (old) at (.75,1.20)
    {\textbf{Old \(C_0\)}\\
     req. \(\rightarrow\) \texttt{None}};
  \node[active,text width=1.55cm,anchor=west] (new) at (2.60,1.20)
    {\textbf{Turn 2: \(C^\star\)}\\
     req. \(\rightarrow\) raise\\
     opt. \(\rightarrow\) default};
  \draw[->,specRed,line width=.8pt] (old.east)--(new.west);

  \path[rounded corners=3pt,draw=specBlue,line width=.8pt,fill=specBlueLight]
    (4.75,.45) rectangle (7.95,4.15);
  \node[resolve,text width=2.45cm] (rd) at (6.35,3.35)
    {\(\operatorname{Resolve}(H_D)=C^\star\)};
  \node[align=center,font=\scriptsize,text=specBlue] at (6.35,2.27)
    {\textbf{same active atoms}\\
     required: raise \(\cdot\) optional: default};
  \node[resolve,text width=2.45cm] (ro) at (6.35,1.20)
    {\(\operatorname{Resolve}(H_O)=C^\star\)};

  \path[rounded corners=3pt,draw=specGray!60,line width=.7pt,fill=specGrayLight]
    (8.55,.45) rectangle (12.10,4.15);
  \node[align=center,font=\tiny\bfseries,text=specGray] at (10.325,2.42)
    {SAME CONTROLS FOR BOTH RUNS};
  \node[align=center,font=\tiny,text=specGray,text width=3.05cm]
    at (10.325,2.08)
    {repo \(\cdot\) agent configuration\\20-min budget \(\cdot\) verifier};
  \node[run,text width=2.55cm,anchor=west] (runD) at (8.90,3.03)
    {\textbf{Agent run D}\quad fresh \(R^0\)\\[2pt]
     resulting patch \(P_D\)};
  \node[run,text width=2.55cm,anchor=west] (runO) at (8.90,1.20)
    {\textbf{Agent run O}\quad fresh \(R^0\)\\[2pt]
     resulting patch \(P_O\)};

  \node[result,text width=3.55cm,anchor=west] (outD) at (12.75,3.03)
    {\textbf{Required-missing probe}\\[2pt]
     raises \texttt{SecretNotFoundError}
     \hfill\textcolor{specTeal}{\textbf{\checkmark}}\\
     \textcolor{specGray}{Block score}\hfill\textbf{5/5}};
  \node[result,text width=3.55cm,anchor=west,draw=specRed!80,
    fill=specRedLight] (outO) at (12.75,1.20)
    {\textbf{Required-missing probe}\\[2pt]
     returns \texttt{None}
     \hfill\textcolor{specRed}{\textbf{\(\times\)}}\\
     \textcolor{specGray}{Block score}\hfill\textbf{4/5}};
  \node[anchor=north east,font=\tiny\bfseries,text=specRed] (stale)
    at (16.82,.52) {stale behavior from superseded \(C_0\)};
  \draw[->,specRed,line width=.75pt] (stale.north) -- (outO.south east);

  \draw[flow] (direct.east)--(4.75,3.35);
  \draw[flow] (new.east)--(4.75,1.20);
  \draw[flow] (7.95,3.35)--(runD.west);
  \draw[flow] (7.95,1.20)--(runO.west);
  \draw[flow] (runD.east)--(outD.west);
  \draw[flow] (runO.east)--(outO.west);
\end{tikzpicture}}
    \caption{A specification-path counterfactual from Tracecat PR~\#1245.
    The direct and override histories resolve to the same active contract
    $C^\star$, while repository, agent configuration, budget, and verifier are
    held fixed across fresh runs. Nevertheless, the override run preserves the
    superseded behavior and fails the required-missing probe. \benchmark\ scores
    this verified behavioral difference rather than patch identity.}
    \label{fig:overview}
\end{figure*}
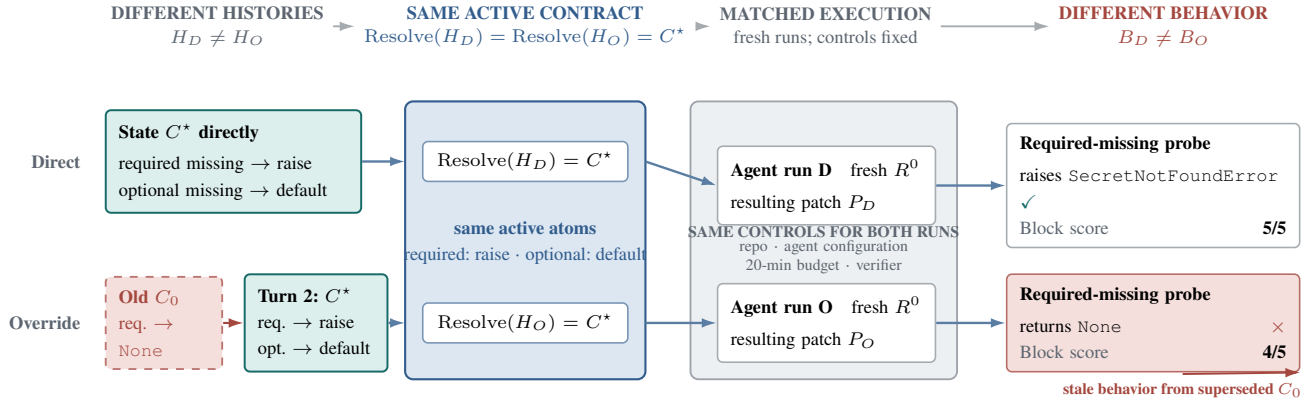

Figure~\ref{fig:overview} shows the problem using a real task derived from
Tracecat PR~\#1245. The final contract distinguishes required from optional
secret retrieval. A missing required secret must raise
\texttt{SecretNotFoundError}; an optional lookup must return a supplied default
or \texttt{None}. The direct history states both rules once. The override
history first asks for the old behavior and then explicitly replaces it with
the same two rules. Thus the histories differ in route but not in what remains
binding.

Canonical repository benchmarks usually expose one consolidated issue and ask
whether an agent can implement it
\citep{jimenez-2024-swebench,yang-2024-sweagent,xia-2025-demystifying,searchgen}.
Multi-turn and evolving-code benchmarks reveal interaction, accumulation, and
long-context failures, but often change the active target or workspace together
with the conversation
\citep{rawal-2025-mtsec,wang-2026-codeflowbench,
yan-2026-emergent-specification,raghavendra-2026-sweinteract,
shen-2026-evocodebench,deng-2026-swemilestone,badseeing}. In those settings, presentation,
the active target, and accumulated code can change together. They therefore do
not provide the counterfactual in Fig.~\ref{fig:overview}: when only the path to
a fixed final contract changes, does the same agent system remain
contract-correct?

We call the observed dependence \emph{specification-path sensitivity} and the
desired robustness property \emph{specification-path invariance}. If histories
$H$ and $H'$ resolve to the same final contract $C^\star$, a path-invariant
agent should remain contract-correct under a fixed repository, verifier, and
execution policy. Patch or trace identity is unnecessary because many programs
may satisfy the same behavioral obligations. What should remain invariant is
tested contract realization.

Measuring this property requires more than splitting a prompt into several
turns. First, equivalent histories must agree on every active obligation,
including scope and polarity; topical similarity is not enough. Second,
revision semantics must be distinguished from ordinary effects of wording,
repetition, length, and integration. Third, the comparison must preserve the
repository and execution policy so that the history itself is the manipulated
factor. Finally, correctness must be behavioral rather than textual: two
different patches may both be correct, while a plausible-looking patch may
retain an explicitly superseded rule.

\benchmark\ turns these requirements into a controlled diagnostic evaluation.
For each task family, it constructs multiple histories that resolve to one
final contract, then holds the repository, verifier, agent configuration, and
execution budget fixed across histories. Direct and control conditions separate
canonical implementation competence from sensitivity to revision path, wording,
and added context. Each execution begins only after the complete history is
visible and starts from a fresh copy of the same base repository. This deferred
execution design isolates contract resolution from the inertia of code written
at intermediate turns.

The central empirical result is a failure of conditional robustness, not a
leaderboard-wide accuracy drop. Final-contract realization (FCR) is
\directfcr\ for the direct condition and \pathfcr\ when averaged over the four
contract-equivalent history conditions. Yet 35 of 100 complete blocks that
succeed directly fail on at least one alternative. Thus average success can
remain stable while the identities of successful executions change across
equivalent paths.

This work makes three contributions:
\begin{itemize}
    \item We identify specification-path sensitivity and formalize the
    corresponding invariance criterion over contract-equivalent requirement
    histories.
    \item We introduce \benchmark, which isolates revision path while holding
    the repository, final contract, verifier, agent configuration, and budget
    fixed.
    \item We show that stable aggregate accuracy can conceal substantial
    path-conditioned failures among executions that succeed on the direct
    specification.
\end{itemize}

\section{Related Work}
\label{sec:related-work}

\subsection{Repository-Level Coding-Agent Evaluation}

Repository benchmarks pair issues with code snapshots and executable tests;
agent studies add repository navigation, editing, and validation
\citep{jimenez-2024-swebench,yang-2024-sweagent,
xia-2025-demystifying,rationalrewards}. RepoBench instead studies repository-aware completion
\citep{liu-2024-repobench}. These settings establish realistic implementation
competence, but normally present one consolidated specification. They do not
test whether the same contract is recovered from alternative histories.

\subsection{Interactive Coding and Evolving Requirements}

Fixed-goal conversations, alternative construction orders, and progressively
disclosed tasks show that presentation alone can change coding performance
\citep{rawal-2025-mtsec,wang-2026-codeflowbench,
yan-2026-emergent-specification,raghavendra-2026-sweinteract,
laban-2026-lost-conversation,emergent}. Benchmarks of genuinely evolving software also
expose adaptation, stale behavior, and regression
\citep{wang-2025-codeifbench,zhan-2025-sreval,
wang-2025-maintaincoder,sobal-2026-staminabench,
shen-2026-evocodebench,deng-2026-swemilestone,
lam-2026-swechain,huang-2026-regression,wang2025code}. These are essential precedents, but
their active target, workspace, or prior implementation often changes with the
conversation. \benchmark\ holds those factors fixed and varies complete
contract-equivalent histories.

\subsection{Instruction Updating and History Sensitivity}

Instruction-following work studies replacement, prioritization, retraction, and
stale state across turns
\citep{rakotonirina-2025-teammates,han-2025-entangled-instructions,
yan-2024-refutebench,zhu-2026-aging-agents,zhai-2026-revisable,reverse}.
Long-context results show that length, position, splitting, and interference
are competing explanations for any history effect
\citep{bai-2024-longbench,liu-2024-lost-middle,
laban-2026-lost-conversation,gupta-2024-interference,jaffe-2026-rift}.
Accordingly, duplicate, split, paraphrase-direct, and length-matched conditions
help separate revision structure from repetition, integration, and generic
context burden. They do not by themselves identify an internal mechanism.

\subsection{Metamorphic and Executable Evaluation}

Metamorphic testing evaluates relations between transformed inputs and outputs
\citep{segura-2016-metamorphic}; behavioral checklists and contrast sets turn
this idea into matched counterfactual tests
\citep{ribeiro-2020-checklist,gardner-2020-contrastsets}. Code-generation work
has applied semantics-preserving prompt and history transformations
\citep{wang-2023-recode,wang-2024-metamorphicprompt,guo-2026-mortar}.
\benchmark's input relation is equality of the resolved final contract; its
output relation is contract-correct executable behavior, not patch identity.
Because tests are incomplete oracles, claims remain bounded by tested behavior
\citep{barr-2015-oracle,liu-2023-evalplus,
wang-2026-swebenchcorrectness}. Requirements-change taxonomies inform the
history operators and validation gates
\citep{madampe-2022-requirements-changes,zowghi-2003-threecs}.

\paragraph{Positioning.}
\benchmark\ contributes the controlled conjunction of these ideas:
independently validated contract-equivalent histories, repository-level paired
execution under a fixed verifier and budget, and conditional measurement among
blocks with demonstrated direct competence. This conjunction makes the path to
the contract observable as a factor rather than leaving it entangled with a
changing goal or workspace.

\section{Problem Formulation}
\label{sec:formulation}

\subsection{Requirements, Contracts, and Histories}

For task $i$, a requirement atom $q\in\mathcal{Q}_i$ is the smallest behavior
that can be activated or revoked and tested independently. It records scope,
input condition, polarity, and required observation. Polarity matters: the
absence of a request is not automatically a prohibition, whereas an explicit
restoration requirement can make retention of an obsolete behavior testably
wrong. Let $M_t$ be the history-manipulated state and $P_i$ fixed obligations
that remain active in every history. The final contract is
\begin{equation}
    C_i^\star = M_i^\star \cup P_i .
    \label{eq:complete-contract}
\end{equation}

Construction uses a hidden trace $Z_H=((A_t,D_t,R_t))_{t=1}^{T}$ of activated,
deactivated, and restated atoms. Starting from $M_0$, replay applies
\begin{equation}
    M_t = (M_{t-1}\setminus D_t)\cup A_t .
    \label{eq:state-update}
\end{equation}
A restatement leaves $M_t$ unchanged; an override deactivates an old atom and
activates its replacement; a cancellation deactivates a temporary atom. The
trace is rendered as natural-language history $H=(u_1,\ldots,u_T)$ without
event labels or atom identifiers.

Histories $H$ and $H'$ are \emph{contract-equivalent} when replay and
normalization give
\begin{equation}
 \mathsf{Norm}(\mathsf{Replay}(Z_H)\cup P_i)
 =\mathsf{Norm}(\mathsf{Replay}(Z_{H'})\cup P_i)=C_i^\star .
 \label{eq:history-equivalence}
\end{equation}
Normalization removes presentation order while preserving atom identity,
scope, polarity, and observation. Because replay validates the hidden program,
not its prose, independent reviewers must also recover the same final contract
from each visible history.

Figure~\ref{fig:overview} instantiates this definition. The direct Tracecat
history activates the two final atoms at once. The override history first
activates an old atom, then deactivates it while activating the same two final
atoms. Replay therefore yields an identical $C^\star$ even though the visible
paths differ. This example also shows why the final turn alone is insufficient:
in a split history, an earlier atom may remain active without being repeated in
the last message.

\subsection{Measurement Unit and Claim Boundary}

A task family $\mathcal{T}_i$ contains one initial repository $R_i^0$, final
contract $C_i^\star$, contract-equivalent histories $\mathcal{H}_i$, verifier
$V_i$, and execution protocol $\Pi_i$. Every history receives a fresh $R_i^0$
under the same $V_i$ and $\Pi_i$. Let $\mathcal{A}(R_i^0,H;\Pi_i)$ be the
repository produced by an agent system. If $V_i$ contains $m_i$ executable
probes, the behavioral signature on repeat $k$ is
\begin{equation}
 S_{i,H,k}=(v_{i1},\ldots,v_{im_i}),\qquad v_{ij}\in\{0,1\}.
 \label{eq:signature}
\end{equation}
The primary outcome requires every final-contract probe to pass:
\begin{equation}
 Y_{i,H,k}=\prod_{j=1}^{m_i} v_{ij}.
 \label{eq:contract-outcome}
\end{equation}

The signature retains more information than the binary outcome. For example,
the illustrated Tracecat block changes from $(1,1,1,1,1)$ under direct to a
signature with one required-lookup failure under override. We use such vectors
to show what behavior changed, while primary inference remains on final-contract
realization. We do not infer an agent's internal representation from a failed
probe.

\subsection{Why Paired Measurement Is Necessary}

Two complementary questions arise once the unit is a matched block. First,
\emph{how often does each condition succeed on average?} This is ordinary
benchmark accuracy and describes end performance. Second, \emph{when direct
succeeds, does the same block remain correct across equivalent histories?} This
is an invariance question. The two answers can diverge because gains and losses
in different blocks cancel. If ten direct successes become failures while ten
direct failures become successes, condition-level accuracy is unchanged even
though the mapping from specification to program is path-sensitive.

Conditioning on direct success serves a specific purpose. A direct failure does
not reveal whether an alternative failure arose from contract resolution or
from inability to implement the consolidated request at all. Direct success
establishes competence for that task, agent configuration, and repeat; failure
elsewhere in the complete block then supplies a counterexample to path
invariance. This diagnostic conditioning does not turn the block into an
independent task or estimate population prevalence.

The task/PR family, not an execution, history, or verifier probe, is the
inferential unit. The five frozen verifiers contain final-contract probes only.
We therefore measure contract realization within their scope; we do not infer
latent beliefs, prove program equivalence, identify stale-state mechanisms, or
estimate the prevalence of revision failures in software tasks generally.

\section{The \benchmark\ Suite}
\label{sec:benchmark}

\subsection{Task Families}

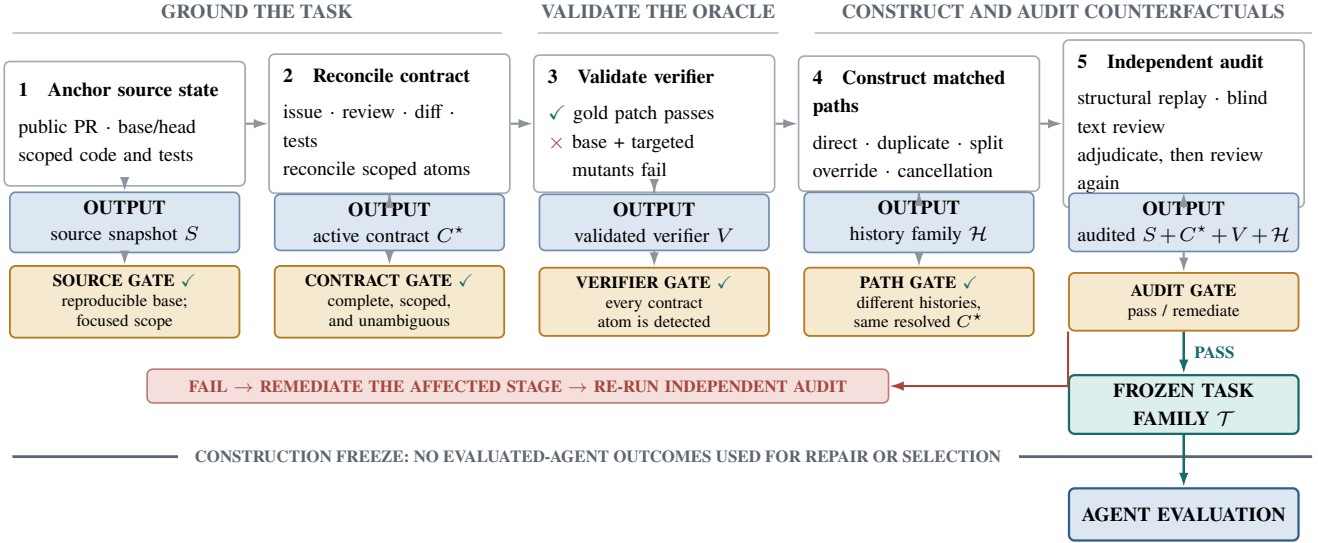
\begin{figure*}[!t]
    \centering
    \resizebox{0.98\textwidth}{!}{\begin{tikzpicture}[
  x=1cm,y=1cm,
  >={Latex[length=1.8mm,width=1.2mm]},
  operation/.style={rounded corners=2pt,draw=specGray!55,line width=.6pt,
    fill=white,text width=2.65cm,minimum height=1.55cm,align=left,
    inner sep=5pt,font=\scriptsize},
  artifact/.style={rounded corners=2pt,draw=specBlue!70,line width=.6pt,
    fill=specBlueLight,text width=2.65cm,minimum height=.62cm,align=center,
    inner sep=3pt,font=\scriptsize},
  gate/.style={rounded corners=2pt,draw=specGold,line width=.65pt,
    fill=specGoldLight,text width=2.65cm,minimum height=.72cm,align=center,
    inner sep=3pt,font=\tiny},
  phase/.style={font=\scriptsize\bfseries,text=specGray,align=center},
  flow/.style={->,draw=specGray!70,line width=.75pt},
  passflow/.style={->,draw=specTeal,line width=.9pt}
]
  \node[phase] at (3.10,5.45) {GROUND THE TASK};
  \draw[specGray!35,line width=.7pt] (.05,5.18)--(6.15,5.18);
  \node[phase] at (8.10,5.45) {VALIDATE THE ORACLE};
  \draw[specGray!35,line width=.7pt] (6.60,5.18)--(9.60,5.18);
  \node[phase] at (13.15,5.45) {CONSTRUCT AND AUDIT COUNTERFACTUALS};
  \draw[specGray!35,line width=.7pt] (10.05,5.18)--(16.25,5.18);

  \node[operation] (s1) at (1.45,4.05)
    {\textbf{1\quad Anchor source state}\\[3pt]
     public PR \(\cdot\) base/head\\
     scoped code and tests};
  \node[artifact] (a1) at (1.45,2.82)
    {\textbf{OUTPUT}\\[-1pt] source snapshot \(S\)};
  \node[gate] (g1) at (1.45,1.83)
    {\textbf{SOURCE GATE \textcolor{specTeal}{\checkmark}}\\
     reproducible base; focused scope};

  \node[operation] (s2) at (4.75,4.05)
    {\textbf{2\quad Reconcile contract}\\[3pt]
     issue \(\cdot\) review \(\cdot\) diff \(\cdot\) tests\\
     reconcile scoped atoms};
  \node[artifact] (a2) at (4.75,2.82)
    {\textbf{OUTPUT}\\[-1pt] active contract \(C^\star\)};
  \node[gate] (g2) at (4.75,1.83)
    {\textbf{CONTRACT GATE \textcolor{specTeal}{\checkmark}}\\
     complete, scoped, and unambiguous};

  \node[operation] (s3) at (8.05,4.05)
    {\textbf{3\quad Validate verifier}\\[3pt]
     \textcolor{specTeal}{\checkmark} gold patch passes\\
     \textcolor{specRed}{\(\times\)} base + targeted\\
     \phantom{\textcolor{specRed}{\(\times\)}} mutants fail};
  \node[artifact] (a3) at (8.05,2.82)
    {\textbf{OUTPUT}\\[-1pt] validated verifier \(V\)};
  \node[gate] (g3) at (8.05,1.83)
    {\textbf{VERIFIER GATE \textcolor{specTeal}{\checkmark}}\\
     every contract atom is detected};

  \node[operation] (s4) at (11.35,4.05)
    {\textbf{4\quad Construct matched paths}\\[3pt]
     direct \(\cdot\) duplicate \(\cdot\) split\\
     override \(\cdot\) cancellation};
  \node[artifact] (a4) at (11.35,2.82)
    {\textbf{OUTPUT}\\[-1pt] history family \(\mathcal H\)};
  \node[gate] (g4) at (11.35,1.83)
    {\textbf{PATH GATE \textcolor{specTeal}{\checkmark}}\\
     different histories, same resolved \(C^\star\)};

  \node[operation] (s5) at (14.65,4.05)
    {\textbf{5\quad Independent audit}\\[3pt]
     structural replay \(\cdot\) blind text review\\
     adjudicate, then review again};
  \node[artifact] (a5) at (14.65,2.82)
    {\textbf{OUTPUT}\\[-1pt] audited \(S+C^\star+V+\mathcal H\)};
  \node[gate] (g5) at (14.65,1.83)
    {\textbf{AUDIT GATE} \quad pass / remediate};

  \foreach \s/\a/\g in {s1/a1/g1,s2/a2/g2,s3/a3/g3,s4/a4/g4,s5/a5/g5}{
    \draw[flow] (\s.south)--(\a.north);
    \draw[flow] (\a.south)--(\g.north);
  }
  \draw[flow] (s1.east)--(s2.west);
  \draw[flow] (s2.east)--(s3.west);
  \draw[flow] (s3.east)--(s4.west);
  \draw[flow] (s4.east)--(s5.west);

  \node[rounded corners=2pt,draw=specTeal,line width=.8pt,fill=specTealLight,
    text width=2.65cm,minimum height=.70cm,align=center,inner sep=3pt,
    font=\scriptsize\bfseries] (frozen) at (14.65,.55)
    {FROZEN TASK FAMILY \(\mathcal T\)};
  \draw[passflow] (g5.south)--node[right,font=\tiny\bfseries,text=specTeal]
    {PASS} (frozen.north);

  \node[rounded corners=2pt,draw=specRed!70,line width=.6pt,
    fill=specRedLight,text width=9.10cm,minimum height=.42cm,align=center,
    inner sep=2pt,font=\tiny\bfseries,text=specRed] (remediate) at (6.35,.78)
    {FAIL \(\rightarrow\) REMEDIATE THE AFFECTED STAGE \(\rightarrow\) RE-RUN INDEPENDENT AUDIT};
  \draw[->,specRed,line width=.75pt] (g5.south west) |- (remediate.east);

  \draw[specGray,line width=1.0pt] (.05,-.10)--(16.25,-.10);
  \node[fill=white,inner xsep=6pt,font=\tiny\bfseries,text=specGray]
    at (7.35,-.10)
    {CONSTRUCTION FREEZE: NO EVALUATED-AGENT OUTCOMES USED FOR REPAIR OR SELECTION};
  \node[rounded corners=2pt,draw=specBlue,line width=.8pt,fill=specBlueLight,
    text width=2.65cm,minimum height=.65cm,align=center,inner sep=3pt,
    font=\scriptsize\bfseries] (evaluation) at (14.65,-.82)
    {AGENT EVALUATION};
  \draw[passflow] (frozen.south)--(evaluation.north);
\end{tikzpicture}}
    \caption{Construction and validation of each \benchmark\ task family.
    Every stage emits a named artifact and must pass its validity gate; a failed
    audit returns the affected stage for remediation and re-review. Only the
    frozen package crosses into agent evaluation, whose outcomes never influence
    task repair or selection.}
    \label{fig:construction}
\end{figure*}

\benchmark\ construction follows five ordered stages: source anchoring,
contract reconciliation, verifier validation, path construction, and independent
audit (Fig.~\ref{fig:construction}). Each stage produces an explicit artifact
and must pass its validity gate before the next stage. This order separates
evidence used to define a task from outcomes later used to evaluate agents. The frozen suite
contains five deliberately curated tasks derived from five public pull requests;
it is a controlled diagnostic suite, not a random sample of software changes.

\subsection{Source Selection and Contract Extraction}

Candidate discovery combined a broad PR funnel with high-touch curation. The
automatic gate retained merged, non-bot changes with both source and test
modifications and excluded changes dominated by generated files or too large
for reliable focused extraction. These signals established constructability,
not semantic ground truth. Final admission additionally required a reproducible
base checkout, an observable behavior change, a focused executable verifier,
and enough evidence to distinguish the final behavior from a plausible
alternative. Task selection did not use scores from the systems reported here.

For each admitted PR, a curator inspected the linked issue, PR description,
review discussion, source diff, and changed tests. The curator then restricted
the task to one coherent behavioral slice and decomposed that slice into atoms
with explicit conditions, scopes, polarities, and observations. This focusing
step prevents a large PR from becoming one opaque label. For example, the
SBSim source PR changes twenty files, but the family retains only bounded
day-of-week and hour-of-day normalization; an adjacent coordinate-regex change
is outside the contract. Similarly, the Tracecat family excludes integration
call sites unrelated to required and optional secret retrieval.

Table~\ref{tab:tasks} summarizes the resulting final contracts. Each family
fixes the base commit, requirement catalog, verifier, and reference patch before
evaluated-agent execution. Extraction models, where used, proposed candidate
observations from the evidence; their proposals were construction aids rather
than labels. Human curation determined the admitted atoms and recorded rejected
or merged candidates.

\begin{table}[t]
    \centering
    \scriptsize
    \setlength{\tabcolsep}{3pt}
    \begin{tabular}{@{}lll@{}}
        \toprule
        \textbf{Repository} & \textbf{PR} & \textbf{Final-contract behavior} \\
        \midrule
        kedro-org/kedro & 4958 & catalog replacement precedence \\
        NVIDIA/NeMo-Agent-Toolkit & 513 & conversion failures raise errors \\
        camptocamp/pytest-odoo & 86 & subtest compatibility restoration \\
        google/sbsim & 113 & bounded time inputs raise errors \\
        TracecatHQ/tracecat & 1245 & required/optional secret retrieval \\
        \bottomrule
    \end{tabular}
    \caption{The five source PR/task families. Descriptions abbreviate the
    executable final contracts; exact atoms, commits, and verifier hashes are
    released with the artifact.}
    \label{tab:tasks}
\end{table}

\subsection{History Construction and Controls}

Every family supplies the five core histories in
Table~\ref{tab:history-families}. Two additional direct conditions are negative
controls. \emph{Paraphrase-direct} is a manually reviewed single-turn
restatement. \emph{Length-matched-no-revision} appends neutral no-op turns to
the direct condition and exactly matches cancellation in turn count and
whitespace-token
count; it is a token proxy, not equality under every provider tokenizer.

The five core conditions change one structural property at a time. Duplicate
tests inert repetition, while split tests whether final atoms introduced in
separate turns are integrated. Override and cancellation introduce the
non-monotonic operations of replacement and retraction. These transformations
do not claim to reproduce the empirical frequency of real review conversations.
They are controlled interventions that make competing explanations measurable.

\begin{table}[t]
    \centering
    \small
    \begin{tabular}{@{}p{0.23\linewidth}p{0.27\linewidth}p{0.40\linewidth}@{}}
        \toprule
        \textbf{Condition} & \textbf{State path} & \textbf{Diagnostic role} \\
        \midrule
        Direct & $M^\star$ & Consolidated-contract competence. \\
        Duplicate & $M^\star\!\rightarrow M^\star$ & Inert repetition. \\
        Split & $\emptyset\!\rightarrow M_1\!\rightarrow M^\star$ & Monotonic integration. \\
        Override & $M_{old}\!\rightarrow M^\star$ & Explicit replacement. \\
        Cancellation & $M_{tmp}\!\rightarrow M^\star$ & Explicit retraction. \\
        Paraphrase-direct & $M^\star$ & Wording control. \\
        Length-matched & $M^\star\!\rightarrow M^\star$ & Extra-history control. \\
        \bottomrule
    \end{tabular}
    \caption{All seven conditions resolve to the same family-specific final
    contract. The first five form the complete block used for CPV.}
    \label{tab:history-families}
\end{table}

\subsection{Executable Calibration and Data Cleaning}

Verifier calibration asks whether each task can distinguish the final contract
from concrete near-misses before it judges an agent. In a fresh base checkout,
the reference patch must pass every final-contract probe and the unmodified base
must fail. Nonempty selective negative patches then remove or reverse one
accepted atom while preserving the others; each must fail its target probe.
Finally, fixed patches are replayed under all five core histories. Their probe
signatures must remain identical across histories, because a verifier whose
behavior changes with prompt presentation cannot support the intended
counterfactual.

This process found and removed several construction artifacts. A pytest-odoo
negative initially materialized as an empty patch and was replaced by a
nonempty statement-level mutant. Parameterized SBSim tests initially produced
incomplete node identifiers, so the curator expanded them to the collected test
nodes and changed the runner to fail closed on empty verifier buckets. A copied
Tracecat package shadowed the patched package on the import path; the duplicate
was removed and the corrected environment was replayed. These are task-cleaning
changes made from verifier evidence, not evaluated-agent performance.

All five accepted families satisfy the same semantic admission content:
provenance is bound to a real PR; atom-to-test coverage is explicit; gold passes
all five histories; base passes none; selective negatives are nonempty and
rejected; patch application is clean; and fixed-patch signature disagreement
across histories is zero. Kedro and pytest-odoo additionally package these
checks in the newer uniform \texttt{construction\_audit\_v4} envelope. The
other three retain earlier family-specific records with the same substantive
checks. We report this artifact-format asymmetry rather than treating the files
as byte-identical evidence.

\subsection{Visible-History Review and Freeze}

Automatic replay first checks that all raw histories end at the same normalized
active set. Two reviewers then independently extract active requirements from
the visible text without the gold trace. Four ambiguous duplicate histories
were revised because a restatement could be read as narrowing the unrepeated
obligations. The revised items were independently re-reviewed rather than
silently relabeled. In the frozen audit, both reviewers exactly match the gold
final set on all 25 histories, agree on all 50 encoded operations, and report no
unresolved ambiguity. The adjudicated gate therefore passes for all five
families.

The full task packages and accepted histories were frozen before the primary
agent matrix. After that boundary, construction records could document a known
asymmetry but could not be upgraded in response to model outcomes. This ordering
is essential: otherwise a benchmark might accidentally clean difficult paths
more aggressively than paths on which agents already succeed.

\section{Experimental Design}
\label{sec:experimental-design}

\subsection{Research Questions and Factorial Matrix}

The experiment follows three questions that mirror the paper's argument.
\textbf{RQ1} asks whether aggregate FCR changes across contract-equivalent
histories. \textbf{RQ2} asks whether direct-success blocks remain correct when
only the path changes. \textbf{RQ3} asks whether wording, added context,
scaffold, missing metadata, or execution time can bound the interpretation of
an observed path violation. RQ1 describes average performance; RQ2 is the
direct invariance test; RQ3 prevents a violation from being assigned to revision
semantics without controls.

We cross five task families, seven model deployments, two scaffolds, seven
conditions, and three repeats:
\begin{equation}
 5\times 7\times 2\times 7\times 3=1{,}470
 \label{eq:factorial}
\end{equation}
planned primary rows. The model deployments are MiniMax-M3, DeepSeek-V4-Flash,
Qwen3.7-Max, Kimi-K2.7-Code, GLM-5.2, GPT-5.4-Mini, and
Grok-4.20-Nonreasoning deployments. The two scaffolds are mini-swe-agent 2.4.5
(\emph{Mini}) and OpenHands CLI 1.16.0 / SDK 1.21.0. We call each
model-deployment--scaffold pair a system configuration; executions are nested
measurements, not 14 independent populations.

Rows are randomized within repeat using seed 20260729, with a barrier between
repeats. All 14 configuration cells passed deployment canaries and synthetic
edit preflights before the primary run. Failed deployments are not silently
substituted.

\subsection{Isolated Execution}

Each row starts an independent session and fresh base workspace. The complete
chronological history is visible before implementation begins. Agent execution
runs in an isolated Podman container; setup and hidden verification run from a
fresh host verifier workspace. The primary agent timeout is 1,200 seconds and
each verifier command has a 120-second timeout. Mini has a USD 0.25 native cost
limit; OpenHands has the same wall-time limit but no equivalent native hard
cost control.

Holding these details fixed is part of the estimand, not merely an engineering
convenience. A scaffold determines how a transcript is rendered, what tools are
available, and how long planning or validation takes. Likewise, a timeout can
turn a nearly complete patch into a fixed-budget failure. We therefore define an
agent configuration by its model deployment, scaffold, prompts, tools, retry
policy, and budget, and compare histories only within that configuration.

\subsection{Run Cleaning and Scoring}

Run processing applies a fixed decision sequence so that infrastructure loss is
not mistaken for agent failure. The scheduler first resolves provider-limit
attempts, selecting the earliest complete non-external attempt under a maximum
of three same-budget tries. The selected record then passes protocol and
metadata gates. Only after those checks do we classify a record as a scored
fixed-budget outcome or as an external invalid. Paired blocks are formed last
and include only histories with scored outcomes.

Agent timeout, no patch, forbidden content, patch-apply failure, nonzero agent
exit, and verifier test failure are fixed-budget failures. Provider, runner,
metadata, protocol, and verifier-infrastructure invalidity are excluded and
reported as attrition. Provider-limit attempts may retry at the same budget up
to three times. This policy selects the first complete non-external attempt and
never changes the primary timeout or model deployment. In particular, an empty
or incorrect patch remains an agent outcome; it is not cleaned away because it
looks uninformative. Conversely, a missing provider response is not scored as
evidence that the agent misunderstood the contract.

\subsection{Sensitivity Analyses}

Two analyses are deliberately separate from the frozen primary estimand.
First, metadata-invalid records with complete verifier evidence are regraded
post hoc; records without that evidence remain excluded. Second, records whose
generation quality was agent-timeout receive one 2,400-second run with the same
task, condition, model deployment, scaffold, and repeat. These reruns join to the
primary schedule by \texttt{parent\_schedule\_id} and are never pooled with the
1,200-second results.

\section{Metrics and Statistical Analysis}
\label{sec:metrics}

\paragraph{Final-contract realization.}
For task $i$, history $h$, and scored executions $\mathcal K_{i,h}$, FCR is
\begin{equation}
 \widehat p_{i,h}=|\mathcal K_{i,h}|^{-1}\sum_{k\in\mathcal K_{i,h}}Y_{i,h,k},
 \qquad
 \widehat P_h=5^{-1}\sum_{i=1}^{5}\widehat p_{i,h}.
 \label{eq:fcr}
\end{equation}
This task-macro quantity describes average success but can hide discordant
outcomes whose net difference is near zero.

\paragraph{Conditional path violation.}
A block $b=(i,e,a,k)$ fixes task, model deployment, scaffold, and repeat. It is
complete when all five core histories are scored. Let $D_b$ denote direct success and let
$R_b$ indicate failure on at least one of duplicate, override, cancellation,
or split. Within task $i$, any-CPV is
\begin{equation}
 \widehat q_i=
 \frac{\sum_{b\in\mathcal B_i}\mathbf{1}[D_b=1]R_b}
      {\sum_{b\in\mathcal B_i}\mathbf{1}[D_b=1]},
 \qquad
 \widehat Q=5^{-1}\sum_{i=1}^{5}\widehat q_i .
 \label{eq:any-cpv}
\end{equation}
Variant-specific CPV replaces $R_b$ with failure under one history. CPV is
undefined when direct fails; such a block is not counted as invariant.

\paragraph{Direct controls.}
For control $c$, we pair scored control and direct-condition rows within the same
task, model deployment, scaffold, and repeat and report the task-macro difference
$\widehat P_c-\widehat P_d$. These controls estimate average wording or
extra-history effects. They are not an any-CPV baseline and therefore cannot,
by themselves, identify revision semantics causally.

\paragraph{Inference.}
Point estimates first average nested observations within each source task and
then weight the five tasks equally. Percentile 95\% intervals use 10,000
nonparametric task/PR-family cluster resamples with seed 20260729. Histories,
repeats, model deployments, and scaffolds never increase cluster $N$. Individual
history-law and deployment comparisons are exploratory; no large-benchmark or
population-prevalence interpretation is made.

\section{Results}
\label{sec:results}

\subsection{Execution Attrition}

Before comparing histories, we separate agent behavior from external invalidity
because missing rows determine which paired blocks are eligible. This accounting
also prevents a provider outage or malformed runner record from becoming an
apparent specification failure.

All 1,470 planned primary rows reached scheduler terminal state. Corrected
selection yields 1,377 completed rows and 93 exhausted provider-limit rows.
The scoring policy retains 1,194 outcomes: 928 final-contract successes and 266
agent failures. The remaining 276 are external invalids: 93 provider, 170
metadata, and 13 verifier invalids (Table~\ref{tab:attrition}). Provider
failures concentrate in GLM and Kimi, with five MiniMax rows; no deployment is
replaced. Because invalidity is uneven across systems, every result reports its
actual support.

\begin{table}[t]
    \centering
    \small
    \begin{tabular}{@{}lr@{}}
        \toprule
        \textbf{Primary-flow quantity} & \textbf{Count} \\
        \midrule
        Planned / terminal & 1,470 / 1,470 \\
        Completed / exhausted provider-limit & 1,377 / 93 \\
        \textbf{Fixed-budget scored} & \textbf{1,194} \\
        \textbf{Final success / agent failure} & \textbf{928 / 266} \\
        External invalid & 276 \\
        \quad provider / metadata / verifier & 93 / 170 / 13 \\
        \bottomrule
    \end{tabular}
    \caption{Primary attrition. External invalid records do not estimate agent
    capability and are excluded rather than scored as failures.}
    \label{tab:attrition}
\end{table}

\subsection{Stable Means Conceal Conditional Path Violations}

The condition-level answer to RQ1 appears reassuring. Task-macro FCR is 78.8\%
(95\% CI 63.6--89.6\%) for direct. Duplicate, override, cancellation, and split
reach 75.8\%, 81.1\%, 80.3\%, and 77.7\%, respectively; their mean is 78.7\%.
The corresponding scored supports are 166, 173, 169, 171, and 165. Thus no
contract-equivalent history produces a large aggregate collapse
(Fig.~\ref{fig:main-results}a).

The paired answer to RQ2 is different. Of 210 possible five-history blocks, 127
have all core histories scored. Direct succeeds in 100 of those complete blocks,
and 35 then fail under at least one contract-equivalent history. The any-CPV
task-macro estimate is 36.4\% (95\% CI 25.6--45.1\%). Improvements on some
blocks therefore offset losses on others: aggregate accuracy remains stable
because it forgets \emph{which} blocks succeed, whereas CPV retains that pairing.

The sensitivity is not confined to one history operator. Duplicate, override,
cancellation, and split produce 19, 8, 13, and 11 positive blocks, with
task-macro CPV estimates of 18.3\%, 10.8\%, 14.1\%, and 12.2\%
(Fig.~\ref{fig:main-results}b--c). Counts overlap because one block may violate
several histories. Duplicate has the largest observed estimate, but the ordering
changes across repeats; individual operator rankings are therefore exploratory.
The result supports path sensitivity as a family of presentation changes, not a
unique failure mechanism tied only to explicit override or cancellation.

\begin{figure*}[t]
\centering
\begin{tikzpicture}[x=1cm,y=1cm]
  \node[anchor=west,font=\small\bfseries,text=specGray] at (0,7.45)
    {(a) Aggregate FCR is nearly stable};
  \begin{scope}[shift={(2.75,4.35)},x=.064cm,y=.40cm]
    \foreach \x/\lab in {0/0,25/25,50/50,75/75,100/100}{
      \draw[specGray!20] (\x,-.35)--(\x,6.35);
      \node[font=\tiny,text=specGray,anchor=north] at (\x,-.55) {\lab\%};
    }
    \draw[dashed,specTeal!70,line width=.7pt] (78.8,-.25)--(78.8,6.25);
    \foreach \y/\name in {6/Direct,5/Duplicate,4/Override,3/Cancellation,
      2/Split,1/Paraphrase,0/Length-matched}{
      \node[font=\scriptsize,anchor=east,xshift=-4pt] at (0,\y) {\name};
    }
    \foreach \lo/\hi/\pt/\y in {63.6/89.6/78.8/6,65.5/85.1/75.8/5,
      67.7/93.2/81.1/4,68.5/88.4/80.3/3,63.0/90.6/77.7/2}{
      \draw[specBlue,line width=.8pt] (\lo,\y)--(\hi,\y);
      \draw[specBlue] (\lo,\y-.13)--(\lo,\y+.13)
        (\hi,\y-.13)--(\hi,\y+.13);
      \fill[specBlue] (\pt,\y) circle[radius=1.7pt];
    }
    \fill[specTeal] (78.8,6) circle[radius=2.5pt];
    \foreach \lo/\hi/\pt/\y in {58.0/88.3/74.2/1,59.4/90.0/74.5/0}{
      \draw[specGold,line width=.8pt] (\lo,\y)--(\hi,\y);
      \draw[specGold] (\lo,\y-.13)--(\lo,\y+.13)
        (\hi,\y-.13)--(\hi,\y+.13);
      \fill[specGold] (\pt,\y) circle[radius=1.7pt];
    }
  \end{scope}

  \node[anchor=west,font=\small\bfseries,text=specGray] at (9.55,7.45)
    {(b) CPV exposes paired failures};
  \begin{scope}[shift={(11.55,4.35)},x=.105cm,y=.56cm]
    \foreach \x/\lab in {0/0,10/10,20/20,30/30,40/40,50/50}{
      \draw[specGray!20] (\x,-.25)--(\x,4.30);
      \node[font=\tiny,text=specGray,anchor=north] at (\x,-.40) {\lab\%};
    }
    \foreach \y/\name in {4/Any,3/Duplicate,2/Override,1/Cancellation,0/Split}{
      \node[font=\scriptsize,anchor=east,xshift=-4pt] at (0,\y) {\name};
    }
    \foreach \lo/\hi/\pt/\y in {25.6/45.1/36.4/4,12.1/25.7/18.3/3,
      3.3/20.9/10.8/2,9.7/18.4/14.1/1,3.8/21.4/12.2/0}{
      \draw[specRed,line width=.9pt] (\lo,\y)--(\hi,\y);
      \draw[specRed] (\lo,\y-.10)--(\lo,\y+.10)
        (\hi,\y-.10)--(\hi,\y+.10);
      \fill[specRed] (\pt,\y) circle[radius=1.8pt];
    }
    \fill[specRed] (36.4,4) circle[radius=2.7pt];
  \end{scope}

  \node[anchor=west,font=\small\bfseries,text=specGray] at (0,3.45)
    {(c) Executable signatures within 100 complete direct-success blocks};
  \begin{scope}[shift={(2.95,.40)},x=.098cm,y=.39cm]
    \foreach \x in {1,...,100}{
      \foreach \y in {0,...,5}{
        \fill[specBlueLight] ({\x-1},\y) rectangle ++(.80,.70);
      }
    }
    \foreach \x in {1,2,3,4,11,12,13,14,36,37,38,39,40,41,42,43,44,
      55,56,57,58,59,60,61,62,79,80,81,82,83,84,85,86,87,88}{
      \fill[specRed] ({\x-1},4) rectangle ++(.80,.70);
    }
    \foreach \x in {1,11,12,13,36,37,38,39,55,56,57,58,79,80,81,82,83,84,85}{
      \fill[specRed] ({\x-1},3) rectangle ++(.80,.70);
    }
    \foreach \x in {1,2,3,36,40,59,79,86}{
      \fill[specRed] ({\x-1},2) rectangle ++(.80,.70);
    }
    \foreach \x in {2,4,11,14,41,42,43,60,61,79,80,81,87}{
      \fill[specRed] ({\x-1},1) rectangle ++(.80,.70);
    }
    \foreach \x in {2,4,44,55,62,79,80,81,82,83,88}{
      \fill[specRed] ({\x-1},0) rectangle ++(.80,.70);
    }
    \foreach \y/\name in {5/Direct,4/Any failure,3/Duplicate,2/Override,
      1/Cancellation,0/Split}{
      \node[font=\scriptsize,anchor=east,xshift=-4pt] at (0,{\y+.35}) {\name};
    }
    \foreach \x in {0,10,35,54,78,100}{
      \draw[white,line width=1.2pt] (\x,-.08)--(\x,5.80);
      \draw[specGray!45,line width=.35pt] (\x,-.08)--(\x,5.80);
    }
    \foreach \x/\name in {5/Kedro,22.5/NeMo,44.5/pytest-odoo,
      66/SBSim,89/Tracecat}{
      \node[font=\tiny\bfseries,text=specGray] at (\x,6.15) {\name};
    }
    \node[font=\tiny,anchor=west] at (103,4.7) {\textcolor{specBlue}{\(\blacksquare\)} pass};
    \node[font=\tiny,anchor=west] at (103,3.8) {\textcolor{specRed}{\(\blacksquare\)} fail};
    \node[font=\tiny,text width=2.3cm,align=left,anchor=north west,text=specGray]
      at (103,3.1) {columns grouped by task and sorted by failure pattern};
  \end{scope}
\end{tikzpicture}
\caption{Stable aggregate accuracy conceals path-conditioned failures. Panel
(a) reports task-macro FCR and 95\% task-cluster bootstrap intervals; the dashed
line marks direct FCR. Panel (b) reports task-macro CPV under the same cluster
bootstrap. In panel (c), each column is one of 100 complete blocks in which
direct passes; cells show whether the same agent configuration realizes the
same final contract under each history. The 35 any-failure columns are a union,
so row counts overlap. Columns are sorted within task only for display. Runs are
nested observations; the five source PR families are the resampling units.}
\label{fig:main-results}
\end{figure*}

The any-CPV signal is present in every source family rather than being driven
by one repository. Positive/eligible counts are 4/10 for Kedro, 4/25 for NeMo
Agent Toolkit, 9/19 for pytest-odoo, 8/24 for SBSim, and 10/22 for Tracecat,
corresponding to within-family rates from 16.0\% to 47.4\%. These five rates are
averaged equally in the 36.4\% macro estimate; the 100 executions do not create
100 inferential clusters. Positive blocks also occur in every repeat. Together,
these checks show that the paired result is replicated across the frozen suite,
while the five-family scale still limits generalization.

\subsection{Scaffold Is a Moderator, Not a Ranked Winner}

Mini contributes 78 complete blocks and OpenHands 49. Their task-macro any-CPV
estimates are 43.0\% (95\% CI 32.9--53.1\%; 57 eligible) and 30.7\% (95\% CI
10.3--51.1\%; 43 eligible), respectively. The intervals overlap substantially,
and OpenHands has wider uncertainty and more invalid support. We therefore
describe scaffold as a plausible moderator, not a ranked winner.

Deployment-specific any-CPV estimates range from 0.0\% to 71.3\%, but eligible
support ranges from only 5 to 28 blocks and three deployment estimates contribute
only four of the five task families. This spread motivates cross-deployment
replication but does not support a stable deployment ranking from the present
data. These comparisons answer whether system components moderate the observed
diagnostic; their uneven support prevents a leaderboard interpretation.

\subsection{Direct Controls Bound the Interpretation}

Relative to the paired direct condition, length-matched-no-revision changes FCR by
$-4.0$ percentage points (95\% CI $-8.1$ to $+0.1$; 151 pairs), while
paraphrase-direct changes it by $-3.1$ points (95\% CI $-10.9$ to $+5.6$; 152
pairs). Both intervals include zero. These controls do not show a
large, precisely estimated average penalty from wording or neutral added
history. However, they are average FCR contrasts, whereas any-CPV is a
one-sided block diagnostic. We therefore conclude that contract-equivalent
presentation can change which executions succeed, but do not attribute the
effect uniquely to non-monotonic revision semantics.

\subsection{Sensitivity Analyses Preserve the Main Boundary}

\paragraph{Metadata regrading.}
Of 170 primary metadata-invalid records, 37 have complete verifier evidence and
can be regraded post hoc; 133 remain excluded. Under this sensitivity, overall
any-CPV changes from 36.4\% to 38.6\% (95\% CI 25.6--48.0\%), and OpenHands
changes from 30.7\% to 34.6\%. The two control contrasts remain near their
primary values. The sensitivity estimate therefore remains close to the primary
estimate, although post-hoc regrading cannot resolve the 133 records without
complete evidence and does not replace primary scoring.

\paragraph{Doubled timeout.}
Additional time recovers many selected timeout cases, showing that budget is
part of the evaluated agent configuration rather than a nuisance variable.
Forty-one generation-timeout records receive the predeclared 2,400-second
sensitivity run, and all finish. Of 32 scored recoveries, 26 succeed and six
remain agent failures; nine further rows are metadata-invalid. Mini recovers
10/10 scored rows and OpenHands 16/22. Because the analysis selects rows by a
primary timeout, it neither retroactively converts the 1,200-second outcomes nor
estimates a population treatment effect.

\section{Discussion}
\label{sec:discussion}

\paragraph{What the central result establishes.}
Canonical success is not sufficient evidence of specification-path invariance.
In this suite, direct and average alternative FCR are almost identical, yet
more than one third of task-macro direct-success blocks fail somewhere else in
the same contract-equivalent family. The logical chain is direct: the direct
condition establishes that a block can realize the consolidated contract; an
alternative history preserves that tested contract; and a subsequent verifier
failure supplies an observed counterexample to invariance for the block. A
condition-level mean discards the pairing and can therefore hide the exchange of
successes and failures.

\paragraph{What it does not establish.}
The experiment does not show that revision histories universally lower coding
accuracy, nor that an internal stale-memory mechanism causes the observed
failures. Duplicate is the largest observed variant-specific CPV, and the two
direct controls have small negative average contrasts whose intervals include
zero. Moreover, separately sampled agent runs remain stochastic even when the
configuration is fixed. The evidence therefore supports a narrower claim:
under matched execution policy, contract-equivalent presentation can change
which directly competent blocks remain correct. The present controls do not
identify a unique internal mechanism or a pure causal effect of non-monotonic
revision semantics.

\subsection{From Long Context to Contract Resolution}

The distinction between retrieval and contract resolution explains why simply
extending context windows does not settle the problem. In the override example,
the obsolete rule remains available; the agent must recognize its supersession
and prevent it from influencing implementation. Split creates the opposite
demand: an earlier atom remains active although the last turn does not repeat
it. Robustness therefore requires selective status tracking, not wholesale
retention or recency.

The observed ordering supports this broader view. Duplicate has the largest
variant-specific estimate, and split also produces violations, so the data do
not isolate non-monotonic updating. Repetition may alter salience and splitting
may alter integration cost. Specification-path sensitivity is the supported
phenomenon; stale memory, recency, and interference remain hypotheses.

\subsection{Implications for Evaluation}

Coding-agent evaluation should report canonical competence and conditional
robustness as separate quantities. FCR answers whether a system commonly
implements a condition. CPV asks whether success on the clearest statement
survives equivalent presentation. Neither subsumes the other: a system may have
high FCR and still expose path violations, while low direct FCR leaves few
eligible CPV blocks. Equal means need not describe the same behavior.

Controls should align with the suspected alternative explanation: paraphrase
for wording, length matching for added context, duplicate for repetition, split
for integration, and override or cancellation for status changes. A compact
report should include direct and condition-level FCR, complete-block coverage,
any- and variant-specific CPV, actual support, and executable signatures. The
raster in Fig.~\ref{fig:main-results} makes visible the overlap that one score
would discard.

\subsection{Implications for Agent Design}

The benchmark suggests a concrete intervention: maintain an explicit contract
ledger before editing. Such a ledger would assign each requirement a stable
identity, scope, polarity, and status such as active, superseded, or canceled.
The implementation plan would then derive from the active subset. This design
targets resolution rather than relying on a summary that may preserve the same
ambiguity in shorter form.

The ledger remains a hypothesis. A valid follow-up should compare it with
token-matched summaries under the same budget and test whether it reduces CPV
without sacrificing direct FCR. An oracle final-contract recap can bound
remaining implementation difficulty, but must not reveal implementation
locations or the reference patch.

\paragraph{Study scale.}
The 1,470 executions replicate configurations and histories, but remain nested
observations of five source tasks. This is a controlled multi-system experiment
and benchmark-design demonstration, not a large task benchmark.

\section{Threats to Validity}
\label{sec:limitations}

\paragraph{External and statistical validity.}
Inference has only five task/PR-family clusters. Cluster-bootstrap intervals
represent variation across these families, not a guarantee of coverage for a
broad software-task population. Deployment-level rankings are especially fragile
because invalidity leaves uneven support; operator comparisons are exploratory.

\paragraph{Missingness and measurement validity.}
Only 127 of 210 possible core-history blocks are complete. Provider failures are
concentrated in particular deployments, and 170 records fail the primary metadata
gate. Attrition tables and post-hoc regrading expose this problem but cannot
prove missing-at-random. Finite final-contract probes may also miss behavior
outside the tested contract. They support tested realization, not program
equivalence or state-level diagnosis. Runs within a block are independently
sampled; CPV is an observed-robustness diagnostic, not a common-randomness
estimate of a deterministic prompt treatment.

\paragraph{Construction validity.}
The visible-history gate passes all five families after documented remediation,
but only two families have the uniform v4 construction matrix. The remaining
human curation and calibration records are accepted under the frozen protocol
but are not treated as identical evidence. Mixed source selection and a focus on
constructable Python PRs prevent a prevalence claim. Synthetic paths provide
control, not their frequency in practice; reviewer disagreement in naturalness
also shows that semantic equivalence need not imply equal conversational style.

\section{Conclusion}
\label{sec:conclusion}

\benchmark\ tests whether coding agents realize one final contract across
equivalent requirement histories. An execution can succeed on a consolidated
request yet fail after revision reaches the same contract, a pattern that
aggregate accuracy hides. Although the mechanism and prevalence remain open,
evaluation should measure specification-path invariance rather than infer it
from canonical success.

\clearpage
\bibliography{specpath_related_work}

\end{document}